\documentstyle[sproclnew,epsfig,12pt]{article}

\def\Journal#1#2#3#4{{#1} {\bf #2}, #3 (#4)}

\def\be{\begin{equation}}
\def\ee{\end{equation}}
\def\bea{\begin{eqnarray}}
\def\eea{\end{eqnarray}}
\def\ltsima{$\; \buildrel < \over \sim \;$}
\def\simlt{\lower.5ex\hbox{\ltsima}}
\def\gtsima{$\; \buildrel > \over \sim \;$}
\def\simgt{\lower.5ex\hbox{\gtsima}}

\begin{document}

\title{TIMING ANALYSIS OF THE SEYFERT-1 GALAXY MCG$-$6-30-15\\WITH XMM-NEWTON DATA}

\author{N. LA PALOMBARA, S. MOLENDI}

\address{IASF/CNR - Sezione di Milano `G.Occhialini'\\Via E. Bassini 15/A, I-20133 Milano (I)\\E-mail: nicola,silvano@mi.iasf.cnr.it}

\author{J. WILMS}

\address{Institut f\"ur Astronomie und Astrophysik - University of T\"ubingen\\Waldh\"auser Strasse 64, D-72076 T\"ubingen (D)\\E-mail: wilms@astro.uni-tuebingen.de}

\author{C. S. REYNOLDS}

\address{Department of Astronomy, University of Maryland\\College Park, MD 20742\\E-mail: chris@astro.umd.edu}

\maketitle\abstracts{\vspace{0.25truecm}We present the main results of a deep timing analysis (in the 0.2-10 keV energy band) of the bright Seyfert-1 galaxy MCG$-$6-30-15 performed on {\em XMM-Newton} data. The light-curves and hardness ratios show that the source has large flux and spectral variations on time-scales of a few kiloseconds at most. From the analysis of both the power spectra and the structure function of the light curves we detected a relevant source variability between time-scales of $\sim$ 0.3 and $\sim$ 10 ks. Based on this result, we extracted a characteristic size of the X-ray emitting region of the order of $10^{13}$ cm.}
\vspace{0.15truecm}

\section{Introduction}

MCG$-$6-30-15 is one of the best studied AGN in the X-ray band, owing to its brightness, strong iron-line and extreme variability. It was observed by {\em XMM-Newton}~\cite{ja} in June 2000, with the {\em EPIC} focal plane camera~\cite{st}$^{,}$~\cite{tu}: in this work we consider the data reported in Wilms et al. (2001)~\cite{wi} from a timing point of view. In this proceeding we restrict our analysis to the {\em pn} dataset.

We extracted a source event list by taking all the data of the {\em pn} camera within an extraction radius of 35$''$ from the source peak: in this way, over 80 \% of the target photons were collected. We selected only single pixel data in order to reject non X-ray events. With an effective observing time of $\sim$ 90 ks (after the rejection of time periods affected by a high proton flux), we obtained a total number of `net' (i.e. background subtracted) counts of 2.85$\cdot10^{5}$, 4.28$\cdot10^{5}$ and 1.44$\cdot10^{5}$ in, respectively, the energy ranges 0.2--0.5, 0.5--2 and 2--10 keV.

\section{Data analysis}

In Fig.1 we show the background-subtracted light curves of the target. All of them display two clear photon flares at about 35 and 95 ks from the observation start; their typical time-scales are of a few kiloseconds. The source flux is far from being constant along the whole observation: at all the energies the light curves show several features. The {\em hardness ratio} HR1 (defined as {\em HR1=cr(0.5-2)/cr(0.2-0.5)}) increases when the source total count rate rises, while HR2 (defined as {\em HR2=cr(2-10)/cr(0.5-2)}) decreases: this indicates that MCG$-$6-30-15 is characterised by spectral variations associated to the intensity variations, with the largest variability in the 0.5-2.0 keV band.

\begin{figure}[t]
\centerline{\epsfig{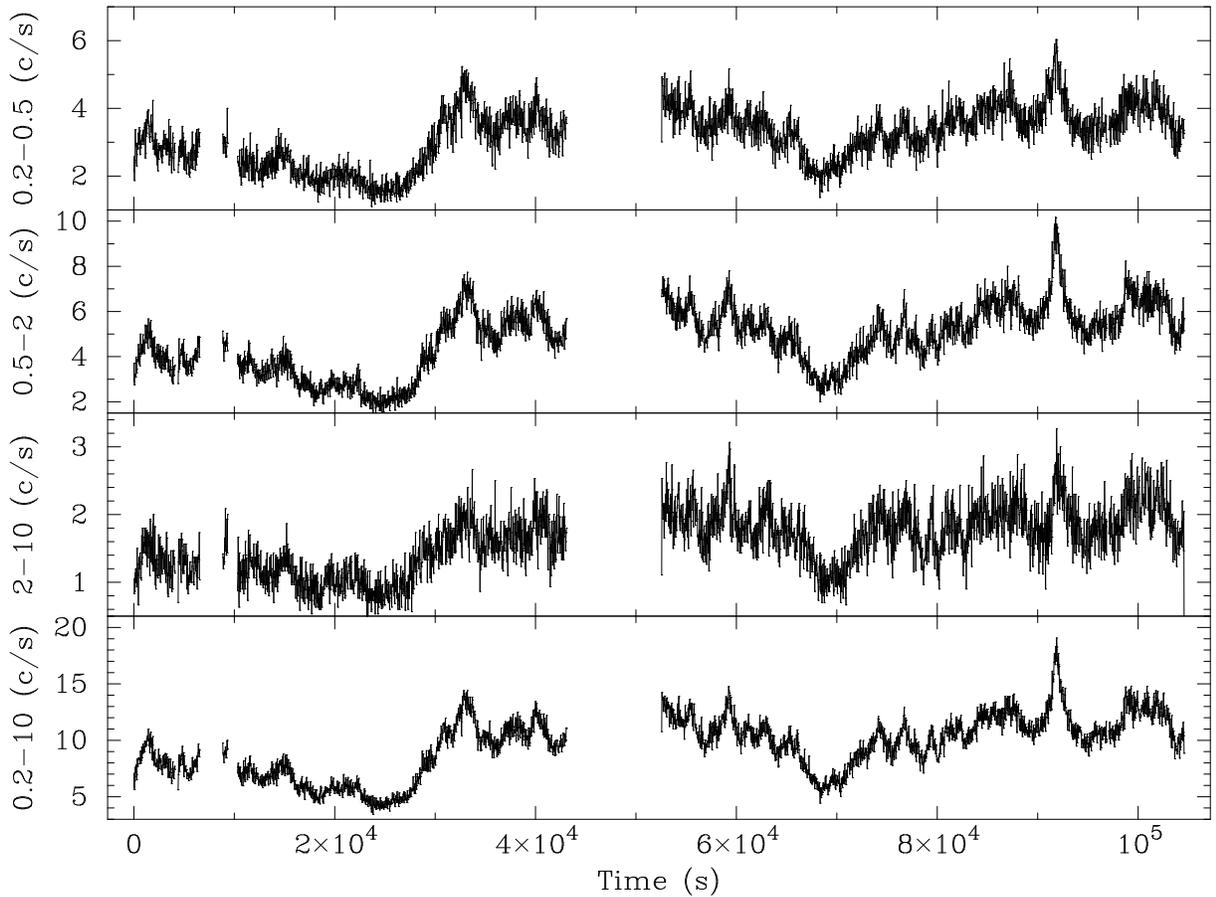}}
\caption{Background-subtracted light curves of the {\em EPIC pn} data of MCG$-$6-30-15 in four different energy ranges}
\end{figure}

In order to characterise the details of the source variability, we applied the technique of the {\em Power Spectral Density} (PSD) analysis to the source data, in the previous four energy ranges. For comparison, we calculated the PSD of four `dummy', stable sources with the same mean count rate and observation length of the corresponding real light curve, also including their proper counting noise: in this way, we can obtain a direct measurement of the expected level of the source Poisson noise. The resulting distributions are shown in Fig.2: there we report, for each energy range, the power spectra of both the target and the stable source and their difference. In this way, thanks to the high throughput of the {\em EPIC} camera, for the first time we can measure the power distribution of MCG$-$6-30-15, with reasonable accuracy, up to frequencies of $\sim 10^{-2}$ Hz.

In all the source PSDs it is possible to identify a characteristic frequency $\nu_{\rm max}\sim$1-3 mHz, which separates two regions with a different appearance: at higher frequencies the spectrum is flat, since it is dominated by the Poisson noise; the opposite is true at lower frequencies, since the intrinsic source power increases, therefore the total spectrum rapidly steepens, down to the minimum sampled frequency $\nu_{\rm min}\simeq10^{-5}$ Hz. Although noise dominates, the noise-subtracted spectra show that some variability can be detected also above $\nu_{\rm max}$, up to the maximum sample frequency of $\simeq$ 10 mHz: but, for all the energy ranges, we can estimate (with a small indetermination) that only a negligible fraction (less than 2$-$3 \%) of the whole background-subtracted source variability is found above 3 mHz. Since the power spectra have a power-law slope of $\sim$2, under the very reasonable assumption that there is no upturn above 10 mHz, we can state that the physical processes responsible for making up  the bulk of the X-ray emission have associated timescales which cannot be smaller than about 1/(a few mHz).

\begin{figure}[t]
\centerline{\epsfig{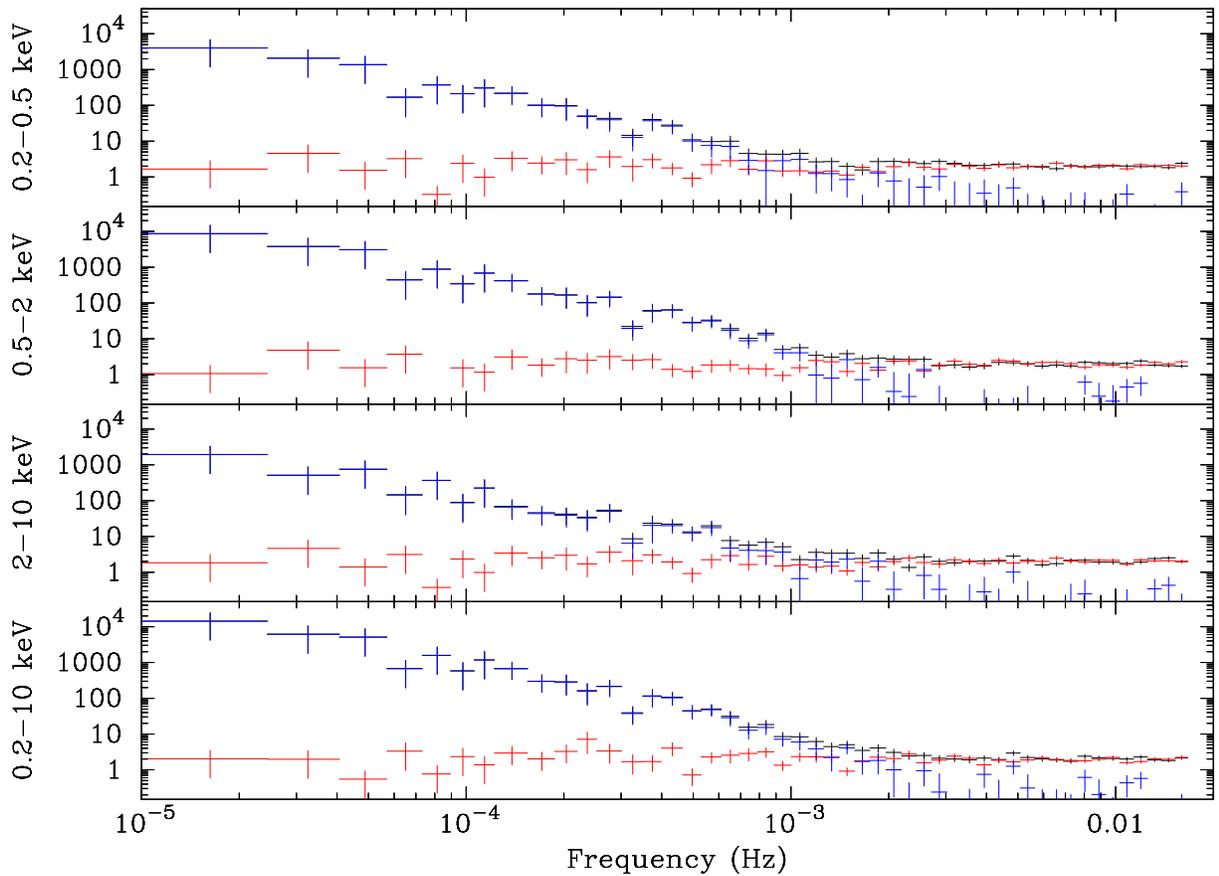}}
\caption{Power Spectral Distribution of the target data (black), a stable source data (red) and the noise-subtracted source data (blue) in four energy ranges}
\end{figure}

As a further step, we calculated the {\em Normalised Structure Functions} (NSFs) of the light curves of MCG$-$6-30-15, in order to characterise its variability time-scales~\cite{zh}: they are shown in Fig.3. They are almost flat for time-scales shorter than $\tau_{\rm min} \sim$ 0.2-0.3 ks; then, they steepen and rise rather smoothly for time-scales between $\tau_{\rm min}$ and a maximum time-scale $\tau_{\rm max}\sim$10-20 ks; eventually, all the NSFs flatten again for $\tau > \tau_{\rm max}$.

\begin{figure}[t]
\centerline{\epsfig{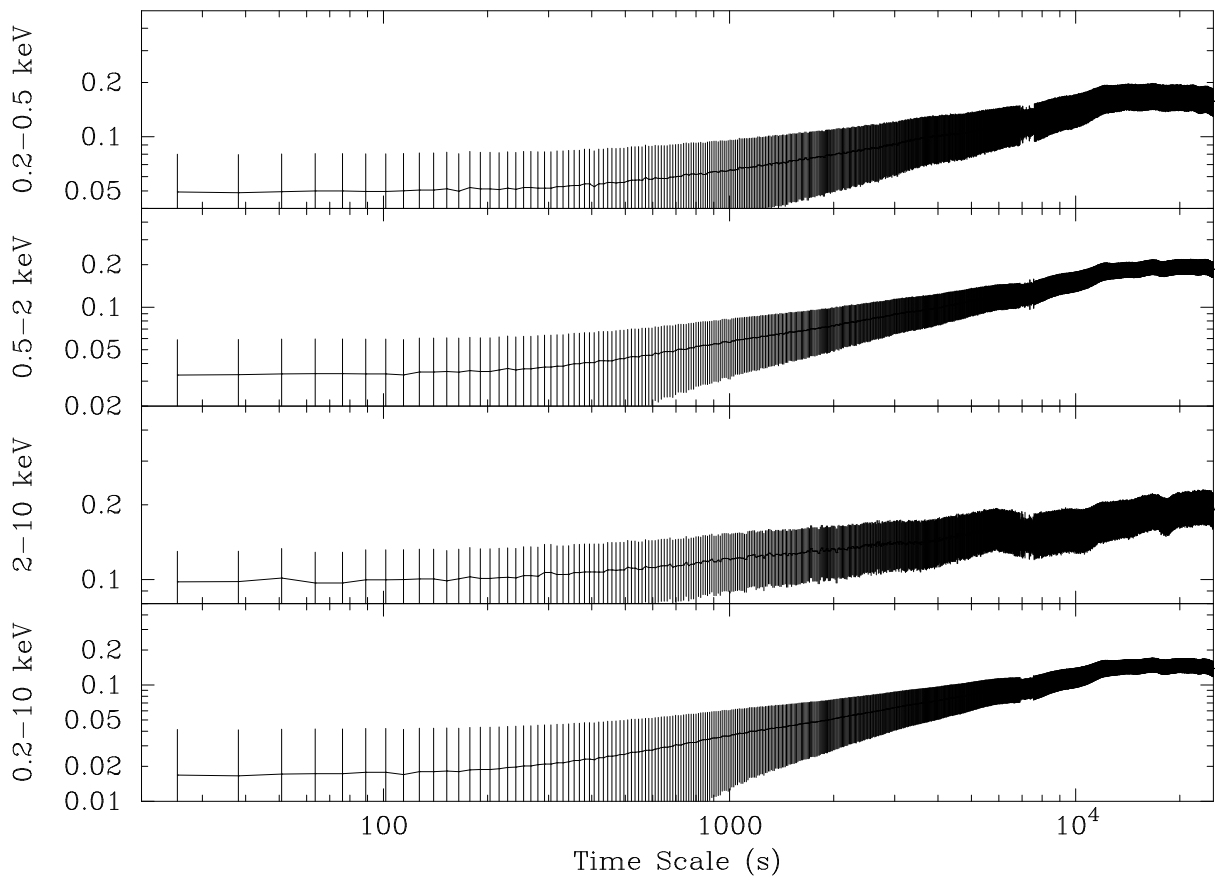}}
\caption{Normalised Structure Function of MCG$-$6-30-15 in four different energy ranges}
\end{figure}

These results suggest that the bulk of the detected source variability is in the range $\tau_{\rm min}$-$\tau_{\rm max}$. It is important to observe that $\tau_{\rm min}\simeq\nu_{\rm max}^{-1}$: thus the {\em PSD} and {\em NSF} techniques both indicates that the measured source variability is dominated by the Poisson noise at $\tau<\tau_{\rm min}$. On the other side, the two techniques provide inconsistent results, i.e. $\tau_{\rm max}<\nu_{\rm min}^{-1}$. Perhaps this result means that our data are unfit to sample the source variability for time-scales $\tau>\tau_{\rm max}$, where we have only a few time-bins.

\section{Summary and conclusions}
The light curves and hardness ratios of MCG$-$6-30-15 show large variations in both its flux and spectrum, with time-scales of a few thousand seconds. The source {\em PSD} proves that there is a meaningful variability power up to $\sim$ 3 mHz; this is confirmed by its {\em NSF}, which also sets an upper limit of $\sim$ 10 ks for the variability time-scale. Based on these results, using the light travel argument we estimate a characteristic scale of $\sim 9 \cdot 10^{12}$ cm for the central source.

The spectral analysis performed by Wilms et al. (2001)~\cite{wi} suggests that the accretion disk extends down to $\sim 2\cdot r_{g} = 6\cdot 10^{13}\cdot m_{8}$ cm. Equating this radius with the one derived above we obtain $M_{BH}\simeq 1.5\cdot 10^{7} M_{\odot}$ and an Eddington luminosity $L_{Edd.} = 1.3\cdot 10^{46}\cdot m_{8}\simeq 2\cdot 10^{45}$ erg/s. The spectral analysis provides a measure of $L_{bol} \simlt 7\cdot 10^{43}$ erg/s, leading to an estimate of $L/L_{Edd.} \simlt$ 3.5 \%.


\section*{References}
\begin{thebibliography}{99}

\bibitem{ja}Jansen F. et al., \Journal{A\&A}{365}{L1}{2001}

\bibitem{st}Str\"uder L. et al., \Journal{A\&A}{365}{L18}{2001}

\bibitem{tu}Turner M.J.L. et al., \Journal{A\&A}{365}{L27}{2001}

\bibitem{wi}Wilms J. et al., \Journal{MNRAS}{328}{L27}{2001}

\bibitem{zh}Zhang Y.H. et al., \Journal{Ap.J}{572}{762}{2002}

\end{thebibliography}
\end{document}